\documentclass{aa}
\input epsf.sty

\begin{document}
\title{Log-parabolic spectra and particle acceleration in the BL Lac object Mkn~421: spectral 
analysis of the complete BeppoSAX wide band X-ray data set}
\author{E.~Massaro\inst{1,2}
\and M.~Perri\inst{1,3}
\and P.~Giommi\inst{3}
\and R.~Nesci\inst{1}
\institute{
Dipartimento di Fisica, Universit\'a La Sapienza, Piazzale A. Moro 2, 
I-00185 Roma, Italy
\and IASF - Sezione di Roma, CNR, via del Fosso del Cavaliere,
I-00113 Roma, Italy
\and ASI Science Data Center, ESRIN, I-00044 Frascati, Italy
}}
\offprints{enrico.massaro@uniroma1.it}
\date{Received 21 May 2003/ Accepted 18 September 2003}

\markboth{E. Massaro et al.: The log-parabolic X-ray spectra of Mkn 421. }
{E. Massaro et al.: The log-parabolic X-ray spectra of Mkn 421. }

\abstract{
We report the results of a new analysis of 13 wide band {\it Beppo}SAX observations
of the BL Lac object Mkn 421. The data from LECS, MECS and PDS,
covering an energy interval from 0.1 to over 100 keV, have been used to
study the spectral variability of this source. We show that the energy 
distributions in different luminosity states can be fitted very well by a log-parabolic
law $F(E)=K~(E/E_1)^{-(a+b~Log(E/E_1))}$, which provides good estimates
of the energy and flux of the synchrotron peak in the SED.  
In the first four short observations of 1997 Mkn~421 was characterized by a very
stable spectral shape, with average values $a=2.25$ and $b=0.45$, independently
of the source brightness and of the fact that the source luminosity was increasing or decreasing. 
In the observations of 1998 smaller values for both parameters, $a\simeq2.07$ and 
$b\simeq$0.34, were found and the peak energy in the SED was in the range 0.5--0.8 keV.
The observations of May 1999 and April--May 2000 covered runs of
a duration of several days and provided a very high number of events for all the
instruments. The resulting spectral fits were then limited by
some instrumental systematics. Also in these cases, the log-parabolic model gave a satisfactory
description of the overall SED of Mkn~421. In particular, in the observations of 
spring 2000 the source was brighter than the other observations and showed a 
large change of the spectral curvature. Spectral parameters estimates gave 
$a\simeq1.8$ and $b\simeq$0.19 and the energy of the maximum in the SED moved to
the range 1-5.5 keV.\\
We give a possible interpretation of the log-parabolic spectral model in terms of
particle acceleration mechanisms. An energy distribution of emitting particles 
with curvature close to the one observed can be explained by a simple model for statistical 
acceleration with the assumption that the probability for a particle to increase
its energy is a decreasing function of the energy itself. 
A consequence of this mechanism is the existence of a linear
relation between the spectral parameters $a$ and $b$, well confirmed by the estimated
values of these two parameters for Mkn~421. 
}
\authorrunning{E. Massaro et al.}
\titlerunning{The log-parabolic X-ray spectra of Mkn 421.}

\maketitle

\section{Introduction}

BL Lac objects (and Blazars in general) are Active Galactic Nuclei (AGN) 
having a polarised and highly variable non-thermal continuum emission extending 
from radio to $\gamma$-rays. Their typical Spectral Energy Distribution (SED) 
is generally described by a double bump structure generated by two emission 
components. 
Current models consider the low frequency bump produced by synchrotron 
radiation (SR) from relativistic electrons in a jet closely aligned to the 
line of sight, and the high frequency bump produced by inverse Compton (IC) 
scattering. 
The peak frequency of the first bump typically ranges from the Infrared to the Optical for the 
so called Low-energy peaked BL Lac (LBL) objects while it is located in the UV-X ray bands for 
the High-energy peaked BL Lac (HBL) (Padovani and Giommi 1995).
To model the shape of these bumps in the $Log(\nu F_{\nu})$ vs $Log~\nu$ plots,
over frequency intervals extending through several decades, different analytical 
laws have been used. These laws are generally based on combinations of power laws,
in some cases with an exponential cutoff, and contain several parameters.
\begin{table*}
\caption{ {\it Beppo}SAX observation log of Mkn 421.}
\label{tab1}
\begin{tabular}{lcccccc}
\hline
 Date & Start UT & End UT & LECS Exp. (s) & MECS Exp. (s) & PDS Exp. (s) & PDS Count Rate$^1$ (s$^{-1}$) \\
\hline
1997/04/29     & 04:02 & 14:42 & 11,562  & ~23,283 & ~18,388 & 0.21 $\pm$ 0.04 \\
1997/04/30     & 03:19 & 14:42 & 11,379  & ~23,975 & ~22,079 & 0.16 $\pm$ 0.06 \\
1997/05/01     & 03:17 & 14:42 & 11,171  & ~23,713 & ~21,748 & 0.13 $\pm$ 0.06 \\
1997/05/02     & 04:10 & 09:41 & ~4,417  & ~11,368 & ~10,594 & 0.31 $\pm$ 0.08 \\
1997/05/03     & 03:24 & 09:41 & ~4,326  & ~11,672 & ~10,734 & 0.10 $\pm$ 0.08 \\
1997/05/04     & 03:25 & 09:45 & ~4,880  & ~12,177 & ~11,168 & 0.16 $\pm$ 0.08 \\
1997/05/05     & 03:32 & 09:45 & ~4,971  & ~11,911 & ~10,946 & $<$ 0.08 \\
               &       &       &         &         &         &  \\
1998/04/21-22  & 01:52 & 03:13 & 23,620  & ~29,547 & ~26,676 &  0.51 $\pm$ 0.05\\
1998/04/23-24  & 00:27 & 06:37 & 27,188  & ~34,656 & ~30,923 &  0.41 $\pm$ 0.05\\
1998/06/22-23  & 07:16 & 02:21 & 11,279  & ~32,516 & ~27,824 &  0.45 $\pm$ 0.05\\
               &       &       &         &         &         &  \\
1999/05/04-08  & 18:16 & 03:21 & 62,868  & 121,813 & 114,693 &  0.14 $\pm$ 0.02\\
               &       &       &         &         &  \\
2000/04/26-/05/03 & 17:36 & 02:13 & 140,374 & ~127,702$^2$ & 151,251 & 3.34 $\pm$ 0.03 \\
2000/05/09-12  & 04:09 & 07:06 & 70,070  & 139,376 & 131,768 & 2.55 $\pm$ 0.02 \\
\hline
\multicolumn{6}{c} { }
\end{tabular}

$^1$ 15--90 keV energy band.\\
$^2$ MECS not operating in the second half of the observation.
\end{table*}

Mkn~421 is a nearby ($z=$0.031) BL Lac object classified as an HBL source because the
energy of the synchrotron peak in the SED is higher than 0.1 keV. 
The spectrum presents a very well marked curvature and the flux changes are 
generally characterized by a high spectral variability (Fossati et al. 2000a). 
The high energy emission of Mkn~421 extends to the TeV range, where it was the
first AGN to be discovered by the Whipple Observatory (Punch et al. 1992).
Mkn 421 has been one of the main targets of multifrequency observational 
campaigns from ground and space observatories. In particular, {\it Beppo}SAX observed 
this source many times, but, because of the large amount of data collected in these 
occasions and the complexity of their analysis, a complete report of all these 
observations has not been published yet.

We present here a new study of the spectral properties of the Mkn~421 SEDs based on 
the entire collection of the {\it Beppo}SAX observations performed with the Narrow Field 
Instruments (NFIs) on board this satellite: LECS (0.1--10 keV) (Parmar et al. 1997), 
MECS (1.3--10 keV) (Boella et al. 1997) and PDS (13--300 keV) (Frontera et al. 1997). 
Some of these observations have already been analysed by other authors while a spectral 
analysis of 1999 and 2000 observations is presented here for the first time.

A time and spectral analysis of the 1997 and April 1998 {\it Beppo}SAX observations of Mkn 421 
was presented by Fossati et al. (2000a, 2000b), while some
results of the observation performed in June 1998 were presented by Malizia et al. 
(2000). 
To fit the spectra these authors used a modified formula of a continuous combination 
of two power laws, including several parameters whose physical interpretation is 
not direct.
Furthermore Fossati et al. (2000b) used in their spectral fits only the LECS and 
MECS data and considered the PDS data separately, while fitting all these data at 
the same time could be more useful for a wide band description of the spectral shape. 

\begin{figure*}
      \vspace{0.5cm}
\epsfysize=9cm
\epsfbox{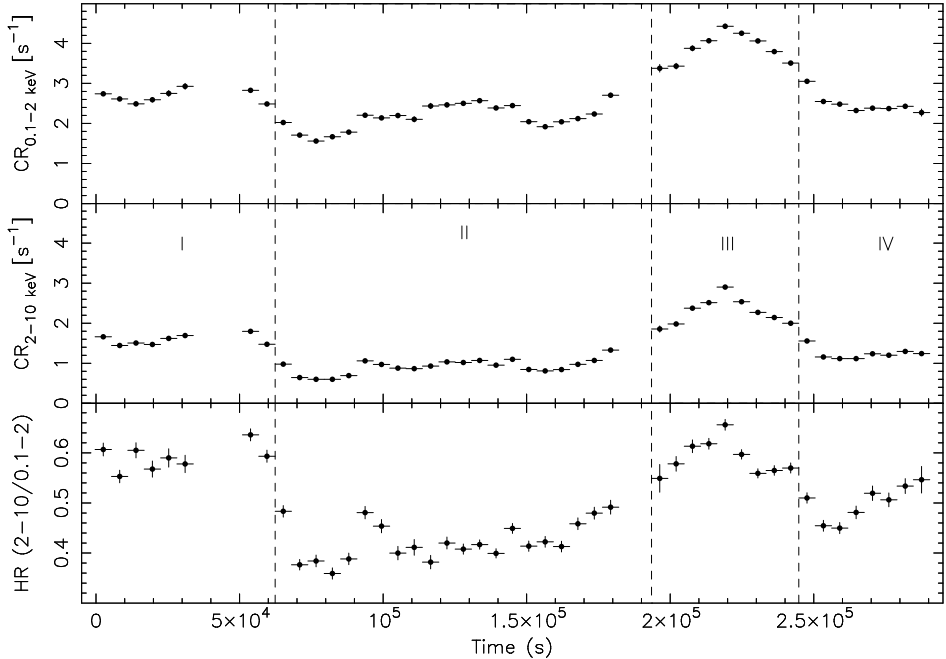}
\caption{The LECS (upper panel) and MECS (central panel) light curves of Mkn 421 during 
the observation of May 4--8, 1999. Data are binned in intervals of 5,700 s, approximately 
corresponding to a {\it Beppo}SAX orbit. The vertical scale is the same in the two panels 
to show the different amplitudes of the intensity variations. In the lower panel the 
corresponding hardness ratio (HR) is plotted.
The dashed vertical lines indicate the four time segments considered in the 
spectral analysis.}
\end{figure*}

In the first part of this paper (Sects. 3 to 5) we show that a simple log-parabolic 
law generally gives a very good description of the X--ray peak in SED of Mkn~421 
also when it is applied to the data of the three NFIs (LECS, MECS and PDS) considered 
together.
The fact that the synchrotron emission of BL Lac objects is generally well described by 
a log-parabolic law was found by Landau et al. (1986) in the analysis of several wide 
band observations (from the millimetric band to UV) of a sample of LBL sources, but these 
authors did not provide a physical explanation, based on the widely accepted emission 
model, for this type of spectral distribution. 
More recently, this spectral model was also used by Krennrich et al. (1999) to 
describe the spectral curvature of Mkn~421 in the TeV range.
In the second part (Sect. 6) of this paper we show that a log-parabolic law is 
naturally expected in the framework of statistical acceleration when a more general 
and simple assumption about the particle probability to increase its energy is 
introduced.
We derive also, from a simple phenomenological approach, the relations between the 
spectral parameters and the general properties of the acceleration mechanism.   
A further advantage of the log-parabolic law is that it can be used to estimate 
several useful quantities, like the peak frequency and bolometric flux of the 
spectral component, in a much simpler way than other models.

During the {\it Beppo}SAX pointing of April 1998, we also performed nearly simultaneous 
optical photometric observations. We use these data to study how the emission in this
frequency range can be linked to the X-ray spectral distribution and show that simple 
extrapolations do not provide acceptable representations of the SED, while a 
log-parabolic model can help to disentangle the possible contributions from different 
emission components.    

Finally we recall that the log-parabolic law has been already used by us to represent
the broad band SED, from the optical to X rays, of some other BL Lac objects like 
OJ~287, MS 1458+22 (Massaro et al. 2003), 3C~66A and ON~325 (Perri et al. 2003). 
We also successfully applied this model in the analysis of the {\it Beppo}SAX X-ray data of Mkn 501 
and the results will be presented in a subsequent paper.

\section{Observations and Data Reduction}

{\it Beppo}SAX observed Mkn 421 on 13 occasions: 7 times in April and May 1997, 
3 times in April and June 1998, 1 time in May 1999 and 2 times in April 
and May 2000. 
In all the observations performed in 1997 the MECS operated with three detectors,
while in those performed in the subsequent years only two detectors were operating.
The logs of all these observations, the net exposure times for the three
NFIs considered in our analysis and the PDS count rates in the 15--90 keV energy band 
are given in Table 1. 
Observations were generally concentrated in time windows of several days: the first 
one had a total duration of ten days from the end of April to the beginning of May 
1997.
The pointings of 1999 and 2000 were much longer than those of the previous periods.
In May 1999 Mkn~421 was observed for about 4 days and a longer uninterrupted run 
was from April 26 to May 3, 2000, followed by another observation started on May 9 
and lasted more than 3 days.
In all the observations the source showed a high variability, characterized by
an irregular oscillating behaviour. The time analysis of the 1997 and April 1998 
data has been presented by Fossati et al. (2000a) and Zhang (2002). 
Malizia et al. (2000)  reported the light curves of June 1998 and those of the 
long observations of April and May 2000 are shown in a paper by Tanihata et al. 
(2002). We do not report in the present paper the 1997 and 1998 light curves and recall
only that Mkn~ 421 was always observed to vary with a count rate amplitude of
a factor of about 2. In some short pointings the flux showed a regular trend: for
instance on 1997 April 29 and May 3 the count rate decreased for the entire duration
of the observation, while on 1997 May 1 it was increasing.  

Standard procedures and selection criteria were applied to the data to 
avoid the South Atlantic Geomagnetic Anomaly, solar, bright Earth and particle 
contamination using the SAXDAS (v. 2.0.0) package.
Data analysis was performed using the software available in the XANADU 
Package (XIMAGE, XRONOS, XSPEC). The images in the LECS and MECS showed 
a bright pointlike source. Events for spectral and timing analysis were 
extracted following the standard procedure currently in use at the {\it Beppo}SAX 
Science Data Center (SDC): data were selected in circular regions, centred at the source 
position, with a radius of 8$'$ for the 1999 and 2000 observations, in which the count rate 
was very high, and 6$'$ and 4$'$ for the LECS and MECS, respectively, in the other 
observations. The response matrices and .arf files used in our analysis have been taken from the 
{\it Beppo}SAX SDC ftp server (September 1997 release) and background spectra were taken from 
the blank field archive.

\begin{figure*}
      \vspace{0.5cm}
\epsfysize=9cm
\epsfbox{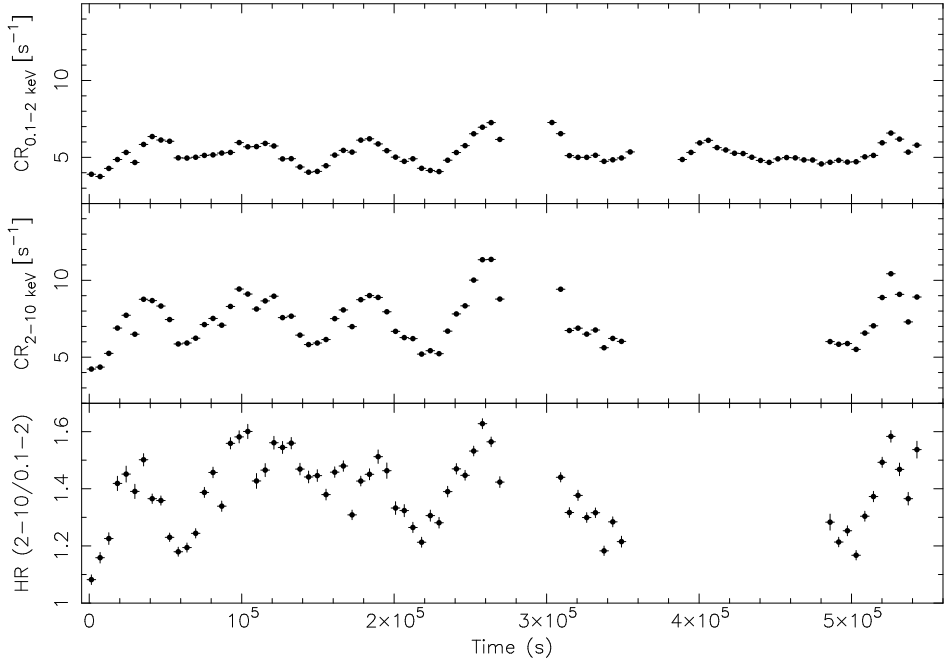}
\caption{The LECS (upper panel) and MECS (central panel) light curves of Mkn 421 during 
the observation of April 26--May 3, 2000. In the lower panel the 
corresponding hardness ratio (HR) is plotted. MECS was not operating in the time interval
from 360,000 to 480,000 seconds after the beginning of the observation. Only the data
taken in the first 360,000 seconds (April 26--30) were used in the spectral analysis. Data are binned 
in intervals of 5,700 s, approximately corresponding to a {\it Beppo}SAX orbit. 
The vertical scale of the two panels are equal to show better the different intensity 
variations.}
\end{figure*}

The light curves in the LECS (0.1--2.0 keV) and MECS (2.0--10 keV) bands of the 
observations of May 1999, April and May 2000 and the corresponding hardness ratios (HR) 
are shown in Figs.~1, 2 and 3, respectively.
The time binning used in these plots corresponds to one {\it Beppo}SAX orbit (about 5,700 s)
and therefore it is not possible to see the hard lags shorter than 3 ks reported by
Fossati et al. (2000a) and Zhang (2002). The count rate changes appear simultaneous
in the two instruments, but the amplitudes are clearly different: the vertical scales 
of the LECS and MECS light curves are the same to make clearer these 
differences. Note that the same modulation appears in the HR light 
curves, confirming this amplitude difference.

In Fig.~1 a strong flare is present in the 
second half of the observation. The rising portion lasted about 6.4$\times$10$^4$ seconds 
and in this time the MECS count rate increased by a factor of 3.5, while that of
the LECS increased by the smaller factor of 2.3; the decay time was only about 20\% shorter,
indicating that there is no large difference in the rising and dimming phases of
the flare.

The different amplitude ratios in the LECS and MECS intensities imply that this
type of flux variations are generally associated with spectral changes, as also shown 
by the HR variations with time, and this makes more complex the spectral analysis.
Similar behaviours are apparent also in the light curves of the two long observations
performed in April--May and May 2000, shown in Figs. 2 and 3. 
In the course of the former observation Mkn~421 showed an oscillating behaviour with
a typical recurrence time of $\sim$ 1 day. Again, the same pattern is present in the light
curve of May 2000 (Fig. 3), but in the last part of the observation two stronger
flares are present with a very fast decay, a factor of 2.9 in $\sim$ 10$^4$ second
in the MECS band.  

\begin{figure*}
      \vspace{0.5cm}
\epsfysize=9cm
\epsfbox{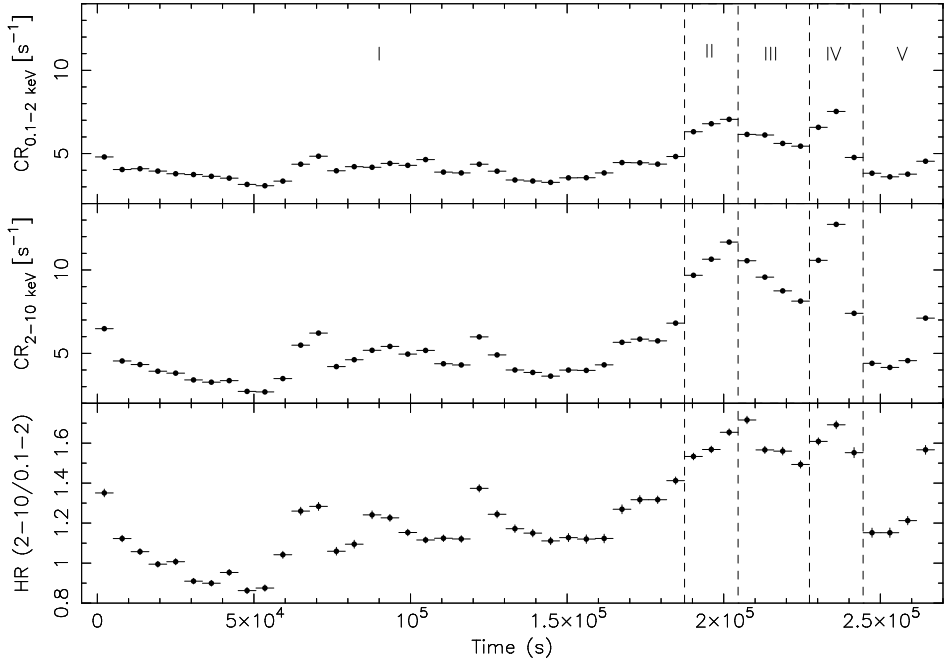}
\caption{The LECS (upper panel) and MECS (central panel) light curves of Mkn 421 during 
the observation of May 9--12, 2000. Data are binned in intervals of 5,700 s, approximately 
corresponding to a {\it Beppo}SAX orbit. The vertical scale is the same in the two panels 
to show the different amplitudes of the intensity variations. In the lower panel the 
corresponding hardness ratio (HR) is plotted. The dashed vertical 
lines indicate the five time segments considered in the spectral analysis.}
\end{figure*}

A relevant point that we want to stress when working with spectra obtained with
two or three NFIs of {\it Beppo}SAX is that of the right choice of the inter-calibration
factors between MECS with LECS ($f_{LM}$) and PDS ($f_{MP}$). The accurate ground
and the in-flight calibrations were used to establish the admissible ranges for
these two factors which are: 0.7$\leq f_{LM}\leq$1.0 and 0.77 $\leq f_{MP}\leq$0.93, 
the latter reduced to 0.86$\pm$0.03 for sources with a PDS count rate higher than 2 ct/s 
(Fiore et al. 1999).
The use of intercalibration factors outside these ranges can affect the evaluation
of the actual wide band spectral distribution, so it is not correct to consider them 
as completely free parameters in the fitting procedure and to fix their values for 
adjusting some results.
In particular, when a spectrum has a well defined intrinsic curvature, like in Mkn~421, 
a not proper choice of the intercalibration factors introduces a bias in the spectral 
estimate. In our analysis we checked accurately that the best values of these two
parameters were in the nominal ranges.

An important problem that we faced in evaluating the best fits originates from
the accuracy of the response matrices and instrumental parameters. 
In the case of the LECS a large feature is
present in the energy range around the carbon edge (0.29 keV), and for long duration observations
with a very high integrated source signal we found statistically significant residuals 
with respect to any smooth model (see Sect.~5 for the case of 1999 observation), already noticed 
by Fossati et al. (2000b) and Malizia et al. (2000). This feature does not change also 
using the 2000 release of the LECS response matrices.
We verified that the same residuals are also detectable in the spectra of other sources 
with a high statistics like Mkn~501 and 3C~273. 
The most likely interpretation is that they are due to instrumental effects like either 
a not very precise evaluation of the LECS effective area in the region of the carbon 
edge or a moderate gain shift. 
In the case of high statistics spectra we have two undesired effects: $i$) the
$\chi^2$ remains high even when the adopted model of the continuum gives a well acceptable
representation of the general spectral behaviour; $ii$) the introduction of a bias in the best fit estimate 
of the spectral parameters.
We verified that after introducing in the XSPEC best fit procedure a moderate gain correction for 
the LECS only, a gain increase of about 3\% minus a bias of only 5 eV, the residuals are very largely 
reduced and much better $\chi^2$ values are obtained (see Sect.~5).

For a correct modelling of the spectral curvature it is very important to known the low 
energy absorption due to the interstellar gas. In our analysis we fixed this  value
to the galactic column density equal to $N_{\rm H}$= 1.61$\times$10$^{20}$ cm$^{-2}$ as 
derived from the 21 cm measure by Lockman \& Savage (1995). One cannot exclude that
a further absorption can be due to the host galaxy. However, high resolution optical
images (Urry et al. 2000) do not give any firm indication for a large absorbing gas
in the galaxian brightness profile and therefore we are confident that the above
$N_{\rm H}$ is well representative of the actual value.

As already shown by Fossati et al. (2000b), the X-ray spectral distributions of Mkn 
421 are remarkably curved. Best fits with simple power laws give largely unacceptable 
$\chi^2$.
One can consider several different analytical models for describing these 
spectra, like a broken power law or a power law with an exponential cut-off 
(PL+EC model) at the energy $E_c$:
\begin{equation}
 F(E) = K~ E^{-\Gamma} {\rm exp}(-E/E_c) ~~~{\rm ph/(cm^2~s~keV)}.  
\end{equation}
This model is generally expected when the spectrum of the emitting electrons has
a sharp or an abrupt upper cut-off in energy because of some limiting processes
in the acceleration mechanisms. It has been considered by several authors and for
this reason we decided to fit it to the data of Mkn~421. As it is shown in sections
4 and 5, the resulting $\chi^2$ are generally too large to be acceptable, indicating
that the actual spectral curvature is milder than an exponential.    
Fossati et al. (2000b) considered a more complex spectral model given by a continuous 
combination of two power laws with photon indices $\Gamma_{-\infty}$ and $\Gamma_{+\infty}$, 
representative of the asymptotic spectral behaviours:
\begin{equation}
 F(E) = K~ E^{-\Gamma_{-\infty}}\left[1+\left(\frac{E}{E_B}\right)^f\right]^{(\Gamma_{-\infty}
-\Gamma_{+\infty})/f} 
~~~,
\end{equation}
without introducing a restriction for their values. In principle it is necessary to 
evaluate the best values of five free parameters. 
Such a law actually provides acceptable fits, but it is not simple to relate the 
parameters' value to the physical quantities of the source model.
Fossati et al. (2000b) did not evaluate the asymptotic spectral indices but 
derived the local slopes at four energies inside the {\it Beppo}SAX band to describe 
better the spectral curvature. They kept the value 
of $f$ fixed to 1 in the analysis of the 1997 data and to 2 in that of 1998, 
without a clear justification based on a physical model of the source.

As stated in the Introduction, in our analysis we preferred to describe the curved 
spectrum of the photon flux with the following log-parabolic (Log-P) model:
\begin{equation}
 F(E) = K~ (E/E_1)^{-(a+b~Log(E/E_1))} ~~~{\rm ph/(cm^2~s~keV)}.  
\end{equation}
We fixed the reference energy $E_1$ to 1 keV and therefore the 
spectrum is completely determined only by the three parameters $K$, $a$ and $b$. 
This model has been already used with success to fit the spectrum of the Crab 
pulsar (Massaro et al. 2000) over several decades in energy and was also applied
in the catalog of 157 X-ray SED of Blazars observed with {\it Beppo}SAX (Giommi et al.
2002).  
The main properties of the log-parabolic are briefly given in the following section.

\section{Main properties of the log-parabolic spectral distribution}

The log-parabolic model of Eq.~(3) is one of the simplest way to represent curved spectra
when they do not show a sharp high energy cut-off like that of an exponential.
It has only one additional parameter with respect to a simple power law and it is also very useful
to estimate other interesting quantities relevant for the physical modelling of
the emission region. 
It is possible to define an energy dependent photon index $\Gamma(E)$, given
by the log-derivative of Eq.~(3):
\begin{equation}
 \Gamma(E) = a~ +~ 2~ b~ Log(E/E_1) ~~~~~~~~.  
\end{equation}
The parameter $a$ is the photon index at the energy $E_1$, while $b$ measures 
the curvature of the parabola: it is easy to demonstrate that the curvature radius
at the parabola vertex is equal to $1/|2b|$. A good estimate of $b$ can be obtained 
only using wide enough energy ranges, particularly when its value is small.

\begin{table}
\caption{ Reduced $\chi^2$ values for the best fit spectra of Mkn~421 with a power 
law with an exponential cutoff and a log-parabola for the observations in 1997 and 1998.}
\label{tab1}
\begin{tabular}{lccc}
\hline
Date & PL+EC & Log-P & d.o.f. \\
\hline
1997-04-29$^{(1)}$ &  2.39 & 1.31 & 132 \\
1997-04-30$^{(1)}$ &  1.82 & 0.75 & 136 \\  
1997-05-01$^{(1)}$ &  2.42 & 1.06 & 132 \\ 
1997-05-02$^{(1)}$ &  1.41 & 0.86 & 127 \\ 
1997-05-03$^{(1)}$ &  1.32 & 0.97 & 100 \\
1997-05-04$^{(1)}$ &  1.55 & 1.16 & 100 \\
1997-05-05$^{(1)}$ &  1.75 & 1.11 & 100 \\  
                   &       &      &     \\
1998-04-21         &  3.73 & 1.19 & 171 \\
1998-04-23         &  2.83 & 1.02 & 169 \\  
1998-06-22         &  2.48 & 1.26 & 150 \\  
\hline
\multicolumn{4}{c} { }
\end{tabular}

(1) ~~PDS data not included.
\end{table}

Another useful form to describe the SED is the following: 
\begin{equation}
 Log(\nu F(\nu)) = Log(\nu_p F(\nu_p)) - b~ \left[Log\left(\frac{\nu}{\nu_p}\right)\right]^2  ~~~~~,  
\end{equation}
where $\nu_p=E_p/h$ is the peak frequency corresponding to the maximum $\nu_p F(\nu_p)$
of the SED.
These quantities can be easily computed from the spectral parameters $a$, $b$ and $K$:
\begin{equation}
 E_p = E_1~ 10^{(2-a)/2b} ~~~~~~~~  
\end{equation}
and 
\begin{eqnarray}
\nu_p F(\nu_p) &=& (1.60\times10^{-9})~K E_1 E_p (E_p/E_1)^{-a/2} ~,  \\
&=& (1.60\times10^{-9})~K E_1^2 ~10^{(2-a)^2/4b} ~~~{\rm erg/(cm^2~s)} \nonumber
\end{eqnarray}
where the numerical constant is simply the energy conversion factor from keV to erg.

This distribution can be analytically integrated over the entire frequency range 
to estimate the bolometric flux. It is not difficult to show that the integral 
of Eq.~(5) between the frequency limits 0 and $\infty$ can be transformed in that 
of a Gaussian function between $-\infty$ and $+\infty$. The final result is:
\begin{equation}
 F_{bol} = \sqrt{\pi~ ln10}~ \frac{\nu_p F(\nu_p)}{\sqrt{b}}~ = 
2.70~ \frac{\nu_p F(\nu_p)}{\sqrt{b}} ~~~~. 
\end{equation}
Considering that typical values of $b$ range around 0.2--0.4 (see the next Section) 
one can obtain a rough estimate of the bolometric flux as 
$F_{bol} \simeq 5~ \nu_p F(\nu_p)$.

\begin{table*}
\caption{ Best fit spectral parameters of the log-parabolic model 
for the 1997-1998 observations of Mkn~421.}
\label{tab1}
\begin{tabular}{cccccccc}
\hline
Date & $a$ & $b$ & $K$ & $E_p$ (keV)  & ~~$\nu_p F(\nu_p)^{(2)}$ & $F_{bol}^{(2)}$ & $F_{2-10~keV}^{(2)}$ \\
\hline
1997-04-29$^{(1)}$ & 2.25 (0.01)   & 0.45 (0.01)    & 7.21 (0.09)~~10$^{-2}$ & 0.53 (.02) & 1.25 (.02) & ~5.0 (.1) &  0.85 (.02)\\
1997-04-30$^{(1)}$ & 2.26 (0.01)   & 0.47 (0.01)    & 7.25 (0.09)~~10$^{-2}$ & 0.53 (.01) & 1.26 (.02) & ~5.0 (.1) &  0.83 (.02)\\
1997-05-01$^{(1)}$ & 2.23 (0.01)   & 0.43 (0.01)    & 8.21 (0.10)~~10$^{-2}$ & 0.54 (.02) & 1.41 (.02) & ~5.8 (.1) &  1.00 (.02)\\
1997-05-02$^{(1)}$ & 2.25 (0.01)   & 0.43 (0.02)    & 1.02 (0.02)~~10$^{-1}$ & 0.51 (.02) & 1.78 (.04) & ~7.3 (.2) &  1.23 (.03)\\
1997-05-03$^{(1)}$ & 2.32 (0.02)   & 0.44 (0.02)    & 6.57 (0.14)~~10$^{-2}$ & 0.43 (.03) & 1.20 (.03) & ~4.9 (.2) &  0.71 (.03)\\
1997-05-04$^{(1)}$ & 2.50 (0.02)   & 0.48 (0.02)    & 4.96 (0.13)~~10$^{-2}$ & 0.30 (.02) & 1.07 (.04) & ~4.2 (.2) &  0.41 (.02)\\
1997-05-05$^{(1)}$ & 2.40 (0.01)   & 0.45 (0.02)    & 6.47 (0.13)~~10$^{-2}$ & 0.36 (.02) & 1.27 (.03) & ~5.1 (.2) &  0.63 (.02)\\
                   &               &                &                        &            &            &           &  \\
1998-04-21~~~      & 2.073 (0.004) & 0.344 (0.006)  & 1.88 (0.01)~~10$^{-1}$ & 0.78 (.01) & 3.04 (.02) & 13.9 (.1) &  3.10 (.02)\\
1998-04-23~~~      & 2.219 (0.004) & 0.373 (0.007)  & 1.35 (0.01)~~10$^{-1}$ & 0.51 (.01) & 2.33 (.02) & 10.3 (.1) &  1.78 (.02)\\
1998-06-22~~~      & 2.066 (0.007) & 0.341 (0.008)  & 1.44 (0.01)~~10$^{-1}$ & 0.80 (.02) & 2.32 (.02) & 10.7 (.1) &  2.39 (.02)\\
\hline
\multicolumn{7}{c} { }
\end{tabular}

Errors are given at 1 sigma for one interesting parameter. \\
(1) ~~PDS data not included. \\
(2) ~~In units of 10$^{-10}$ erg cm$^{-2}$ s$^{-1}$. \\
\end{table*}

Another advantage of the log-parabolic model is that the spectral curvature is 
characterised only by the parameter $b$, while in other models, like the continuous 
combination of two power laws of Eq.~(2), it is function of several parameters. 
A limit of the model, however, is that it can represent only energy distributions 
symmetrically decreasing with respect to the peak frequency. 
More general models, to take into account a possible asymmetry, must  
include at least another parameter. For instance, if at low energies the spectrum 
follows a single power law, while the log-parabolic bending becomes apparent at 
energies higher than a critical value $E_c$, one could use the following mixed model: 
\begin{eqnarray}
F(E)~&=&~K~(E/E_1)^{-(a+b~Log(E_c/E_1))}~~~, E\leq E_c \nonumber \\
F(E)~&=&~K~(E/E_1)^{-(a+b~Log(E/E_1))}~~~,  E > E_c
\end{eqnarray}
however it was not used in the present work since the peak energy of Mkn 421 is 
generally below 1 keV and only limited data are available to detect a significant deviation 
from a symmetric distribution.

\section{The 1997-1998 observations}
\subsection{The X-ray spectral analysis}

In the analysis of the seven observations of spring 1997 we used only the LECS and MECS 
data, while those from PDS were not used in the analysis. In fact, the faint state 
of the source and the spectral steepening at high energies imply a count rate above 15 keV 
much below the sensitivity of the instrument which is limited by confusion due to 
background sources. 
In the three observations of 1998, when the source mean flux was higher than in 
the previous year, PDS events up to the energy of 40--60 keV were also included in the
spectral analysis.

An inspection of the light curves showed that time variability within each pointing was 
limited to about 30\% or less and the hardness ratio was consistent with being constant. 
For this reason spectral fits were evaluated using the events accumulated in the entire duration of
each observation thus allowing us to use the highest statistics for a precise estimation of the 
spectral parameters.

As mentioned in Sect.~2, we fitted the photon spectra using the log-parabolic model 
and obtained statistically acceptable results, much better than those obtained 
with a single or a broken power law. In Table 2 we report the reduced 
$\chi^2$ values and the corresponding d.o.f. for the power law with exponential cutoff
and for the log-parabolic model: it is evident that the former model is generally
not acceptable, while the latter gives reduced $\chi^2$ quite closer to unity.
In particular the power law with exponential cutoff failed to fit the data of the
1998 observations when, because of the higher brightness of the source, there is a
significant detection in the PDS data at energies above 13 keV.

\begin{figure*}
      \vspace{1.0cm}
\epsfysize=9cm
\epsfbox{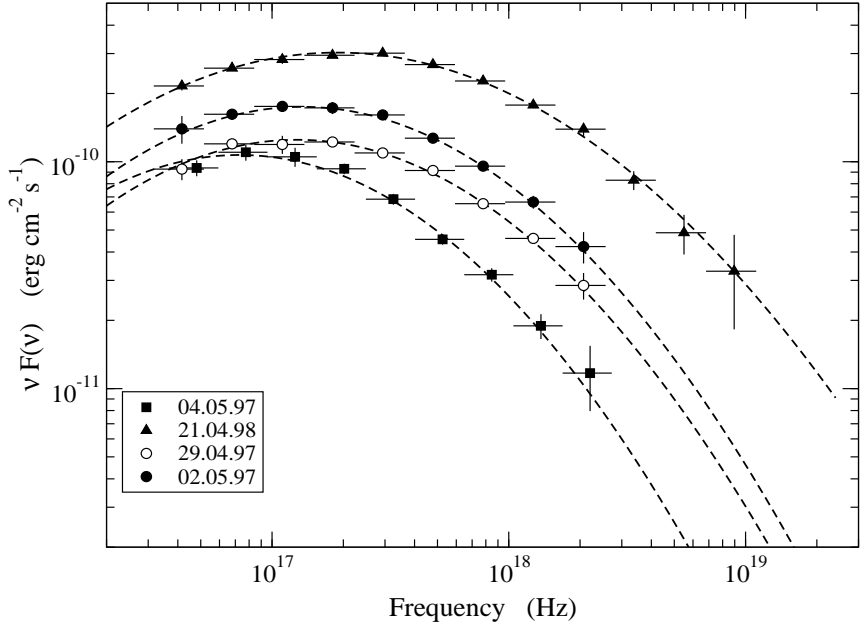}
\caption[]{
The X-ray Spectral Energy Distributions of Mkn 421 in some of the 
{\it Beppo}SAX 1997--1998 observations deconvolved with a log-parabolic law. 
Dashed lines are the best fit to the observed data.
}
\end{figure*}

In Table 3 we report the best fit parameters values for the log-parabolic law, $a$, $b$ and $K$, 
together with three derived parameters, namely the peak energy, the maximum of the SED and the 
bolometric flux of the SR component. Throughout this paper errors are given at 1 standard 
deviation confidence level for one interesting parameter ($\Delta \chi^2 = 1$).
Four examples of these fits at different epochs, including those of the highest and 
lowest intensity are shown in Fig.~4. The best fits of the other observations are 
very similar and they are not shown in the figure for simplicity.

From the data of Table 3 we have important indications about the long term evolution 
of the spectral parameters. During the 7 day campaign of April-May 1997 the spectral 
curvature was remarkably high and very stable, with $b$ in the range 0.43--0.48 with 
a mean value of 0.45. Also $a$, the photon index at 1 keV, changed little: in 
particular, in the first four days it remained practically constant around the mean 
value of 2.25 and increased to 2.32--2.50 in the last three days. 
In the same period the value of the normalisation $K$ instead changed more than a factor of 2, 
indicating that these variations were not accompanied by large spectral changes.
The SED was characterised by a rather low peak energy $E_p$, which ranged in the narrow 
interval 0.30--0.54 keV and by a maximum value $\nu_p F(\nu_p)$ between 1.07 and 1.78 
$\times$10$^{-10}$ erg cm$^{-2}$ s$^{-1}$.
The mean spectral shape remained practically unchanged independently of the fact that
flux was decreasing (1997 April 29, May 3) or increasing (May 1 1997). 

In 1998 Mkn~421 was brighter than in the preceding campaign: the highest count rate
was measured on April 21 and resulted a factor of $\sim$4 higher than in the lowest state
of May 4, 1997. This increase of the total counts had the consequence that the
systematic effects on the LECS spectra, described in Sect. 2, were more relevant than
in the short observations of 1997 and the $\chi^2$ values were slightly higher,
but always acceptable.
Furthermore, we measured a signal in the PDS, up to $\sim$40--60 keV, with
a better evaluation of the curvature parameter because of the more extended range.   
The SED maximum increased to values in the range 2.3--3.0$\times 10^{-10}$ erg cm$^{-2}$ 
s$^{-1}$, but this did not had a clear correspondence in the peak energy. It was 
0.78 keV in the first observation, when the flux from Mkn~421 was high to decrease to 
0.51 after two days when the flux was about 30\% lower.
In the subsequent observation (June 1998), however, when the mean level of the source
was practically the same of April 23, we found an $E_p$ value of 0.80 keV (see Table 3). 
Note also that the $b$ values decreased with respect to 1997, being in the range 
0.34--0.37. This smaller curvature is not due to an enhanced emission at higher 
energies in the PDS range, because it is also well measured using only the MECS data. 
According to Fossati et al. (2000b) on April 23 1998 Mkn~421 was found bright between
12 and 90 keV with a quite hard spectrum: the photon index given by these authors
(2.35$\pm$0.25) is much smaller than that expected from the extrapolation of the 
log-parabolic law with Eq.~(4) and equal to 3.16, despite the large statistical 
uncertainty. 
We verified the count rates in the 12--40 and 40--90 keV ranges and found  
0.318$\pm$0.036 and 0.102$\pm$0.034 respectively, in agreement within the statistical 
errors with those of Fossati et al. (2000b). Again a simple power law fit gives a flat 
spectrum. We note, however, that this count rate level, more than a factor of $\sim$2
higher than that of April 21 when Mkn~421 was brighter, corresponds to a $\nu F(\nu)$
value of $\sim$ 3$\times$10$^{-11}$ erg cm$^{-2}$ s$^{-1}$, moderately larger than the
contribution from the cosmic background. We cannot exclude that this enhanced emission
above 15 keV might be due to some statistical fluctuation, although the interesting
hypothesis that it could be an indication of a new spectral component emerging at higher
energies cannot be excluded.

\begin{figure*}
      \vspace{1.0cm}
\epsfysize=9cm
\epsfbox{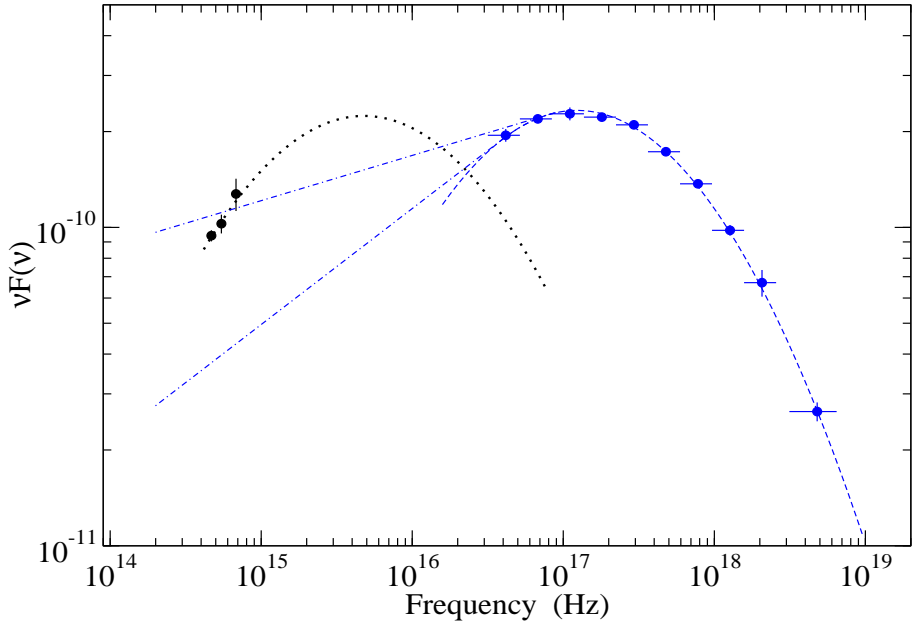}
\caption[]{
The optical to X-ray Spectral Energy Distribution of Mkn 421 observed on 1998,
April 23--24. Two log-parabolic spectral distributions for the optical and X-ray
data sets are shown, together with possible power law extrapolations of X-ray
data at optical frequencies. Notice that these extrapolations cannot match 
the photometric $R$, $V$, $B$ points.
}
\end{figure*}

\begin{figure}
\epsfysize=7cm
\epsfbox{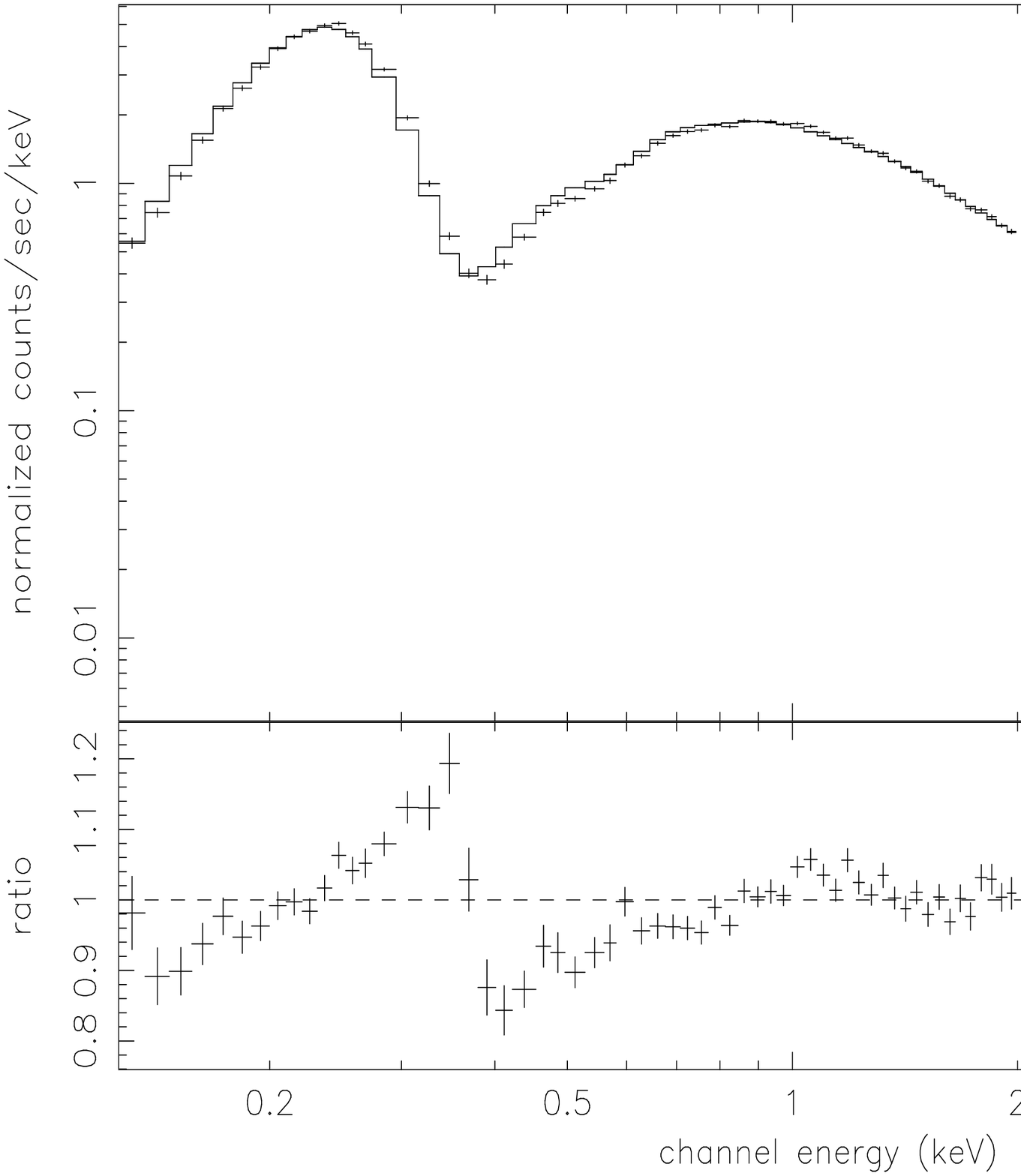}

\epsfysize=7cm
\epsfbox{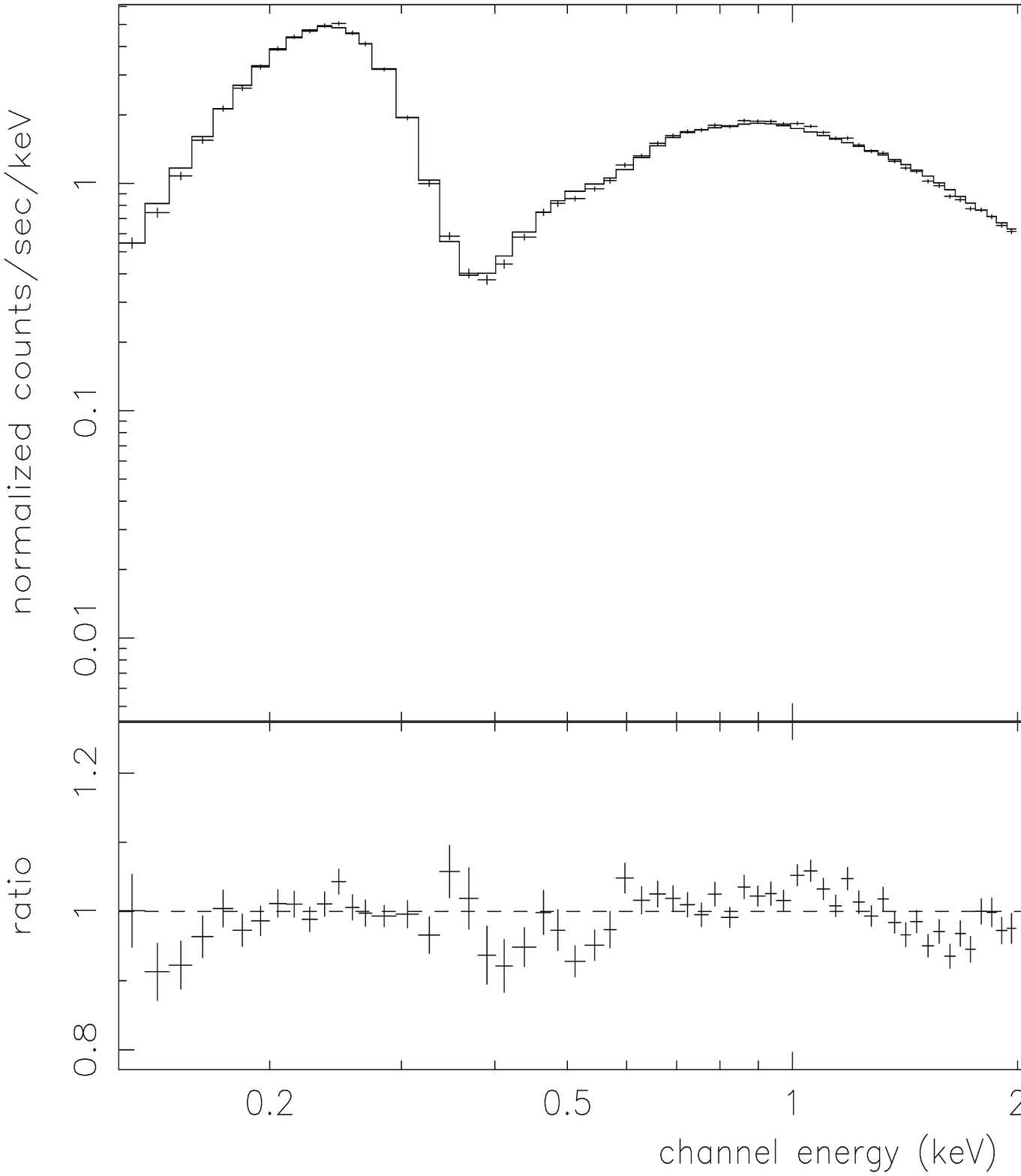}
\caption[]{
Upper panel: best fit of the LECS and MECS data with a log-parabolic law of the 1999 {\it Beppo}SAX 
observation of Mkn 421 without gain correction. Note the large residuals of the LECS points between 0.2 and 
0.6 keV due to some systematics due to the near Carbon edge.

Lower panel: best fit of the same data after the gain adjustment of the LECS. The spurious features at low 
energies practically disappeared.
}
\end{figure}

\subsection{Optical observations}
The low energy tail of the SR peak can extend in the UV--optical range.
Observations at these frequencies are then useful to understand how
wide is the range in which the log-parabolic model gives an acceptable
description of the SED. However, a direct use of photometric data, in
particular if they have been obtained with small aperture telescopes, is
complicated by the fact that the nuclear emission from Mkn~421 is
embedded in that of the bright host elliptical galaxy. To correct the
measured data for the galaxian component it is necessary to know how
much it contributes to the total flux within the used aperture. 
In literature data photometric apertures are not always indicated and 
it is not possible to apply the corrections for the host galaxy contribution 
to derive a reliable estimate of the flux from the nucleus. 

We performed some photometric observations of Mkn 421 in the $B$, $V$
(Johnson) and $R$ (Cousins) bandpasses on April 24, 25 and 28 1998,
just after the {\it Beppo}SAX observations, with the 0.70 cm reflector telescope
of the University of Roma and IASF-CNR located at Monte Porzio and equipped with 
a back-illuminated CCD camera (Site 501A). Standard stars were taken from
Villata et al. (1998). In the three nights, the magnitudes of Mkn 421, 
within a photometric radius of 3.2 arcsec, remained substantially stable 
in a range smaller than 0.1 mag. The measured mean values are:     
$B$=13.26$\pm$0.10, $V$=12.96$\pm$0.05, $R$=12.65$\pm$0.03.
The $V-R$ colour index is in a very good agreement with those
measured in other occasions (see, for instance, Tosti et al.~1998).

We evaluated the host galaxy contribution using the brightness profile
given by Urry et al. (2000), which gives for our photometric radius a
magnitude $R_{gal}$=14.05. The magnitudes in the other two bands were 
computed from the typical colours of an elliptical galaxy (Fukugita et
al. 1995) and resulted equal to $B_{gal}$=15.62 and $V_{gal}$=14.66.
The fluxes of the central component were then computed using the zero
magnitude values from Mead et al. (1990) and obtained $F(R)$=20.1$\pm$0.8,
$F(V)$=18.8$\pm$1.3 and $F(B)$=18.7$\pm$1.9 mJy. 
The optical spectrum of the nuclear source is then quite flat with an energy 
index of 0.3$\pm$0.2. This uncertainty is derived only from the photometric 
errors and it could be even greater if possible systematics, like those derived from 
the galaxy subtraction, are also properly taken into account.

In Fig. 5 we plotted the SED of Mkn 421 derived from these nearly simultaneous
observations. It is clearly apparent that the optical and X-ray data cannot
be described by a single component either with a power law or a log-parabolic
distribution. The extrapolation of log-parabolic best fit of the X-ray data 
into the optical band gives fluxes much lower than the observed ones. Furthermore, 
assuming that the log-parabola could be extrapolated in the optical--UV range by a
power law, as modelled by Eq.~(9), our optical data cannot be fitted since the 
spectral slope is very different from the observed one. Conversely, also the power
law extrapolation of the optical data to the X-ray range does not give the right
fluxes and slope. 

This result suggests that the optical emission does not belong to the same component
of the X rays, although it could still be produced by the same electron population.
In the same Fig. 5 we plotted a possible optical log-parabolic spectrum, 
with the same $b$ value of the X-ray data and the peak value adjusted to have a
small contribution at energies higher than $\sim$ 1 keV: the resulting peak frequency 
would then be around 7.5$\times$10$^{15}$ Hz and the maximum in the SED comparable to 
or slightly higher than that observed in the X rays.
An interesting consequence of this model is that the flux observed at energies lower 
than $\sim$ 0.5 keV is due to the superposition of these two components, which can be
variable with different amplitudes and time scales. This assumption can explain why
the variation amplitudes measured from the LECS light curves are systematically smaller
than those measured from the simultaneous MECS time series (see Figs. 1, 2 and 3).
Finally, we stress that the presence of this low energy component can modify the
spectral distribution producing a rather constant energy index close to unity.
 
\begin{table}
\caption{ Reduced $\chi^2$ values for the best fit spectra of Mkn~421 with a power 
law with exponential cutoff and a log-parabola for the 1999-2000 observations.}
\label{tab2}
\begin{tabular}{lccc}
\hline
Date & PL+EC & Log-P & d.o.f. \\
\hline
1999-05    &  ~4.16 & 1.11 & 141  \\
 ~~~~~I    &  ~1.57 & 1.11 & 111  \\ 
 ~~~~II    &  ~3.22 & 1.14 & 111  \\ 
 ~~~III    &  ~2.51 & 1.06 & 111  \\ 
 ~~~IV     &  ~1.77 & 1.04 & 111  \\ 
           &       &      &      \\
           &       &      &      \\
2000-04    &  13.11 & 1.06 & 152  \\
 ~~~~~H    &  ~8.13 & 1.01 & 152  \\
 ~~~~~F    &  ~8.21 & 1.06 & 152  \\
           &       &      &      \\  
2000-05    &  ~7.89 & 1.34 & 149  \\
 ~~~~~I    &  ~5.82 & 1.37 & 149  \\ 
 ~~~~II    &  ~2.77 & 1.03 & 136  \\ 
 ~~~III    &  ~2.87 & 1.26 & 136  \\
 ~~~IV     &  ~2.14 & 0.87 & 136  \\
 ~~~~V     &  ~1.67 & 1.00 & 121  \\
\hline
\multicolumn{4}{c} { }
\end{tabular}
\end{table}

\section{The 1999-2000 observations}

The three last campaigns in 1999 and 2000 provided long series of data 
with a very great number of events both in the LECS and MECS. 
Because of such high statistics, systematic effects of the detectors, usually 
negligible for other sources, become relevant and produce large $\chi^2$ values which 
can be improved by modifying the instrumental response very little. 
The comparison with a power law plus an exponential cut-off confirmed again that the 
log-parabolic law is better for the general modelling of the continuum. 
Spectral analysis was also applied to data subsets, selected either in particular
time intervals or for different average levels of count rate, to study the possible
relations between $a$ and $b$ and other source parameters.

\begin{figure*}
      \vspace{1.0cm}
\epsfysize=9cm
\epsfbox{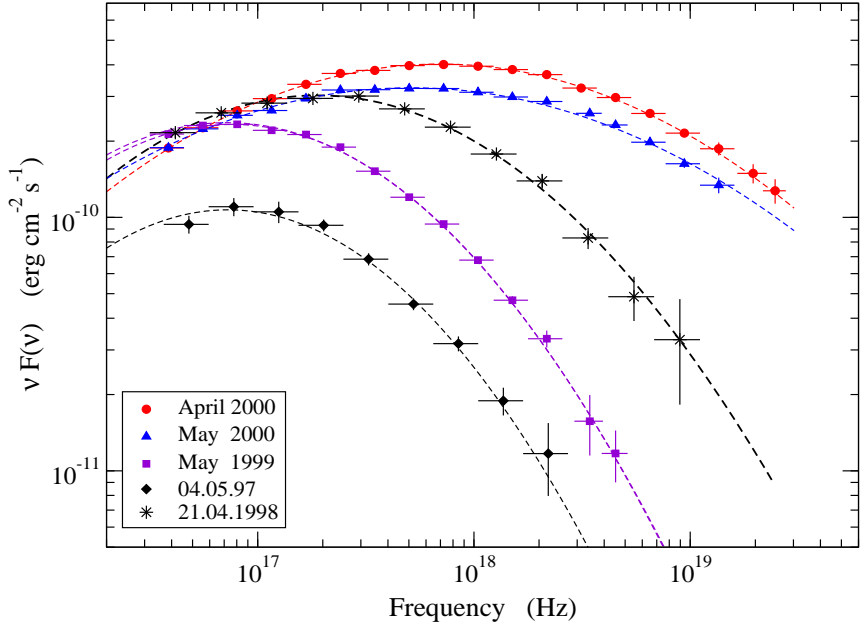}
\caption[]{
The X-ray Spectral Energy Distributions of Mkn 421 in the long {\it Beppo}SAX observations 
of May 1999, April and May 2000. Best fit spectra (dashed lines) were obtained with 
a log-parabolic law. The SED of the two observations of 1997, May 4 and 1998, April 21 are also plotted 
to show the spectral evolution of the source.
}
\end{figure*}
 
The best fit of the four day observation of May 1999 using the log-parabolic model 
applied to the LECS and MECS data gave the unacceptable reduced $\chi^2$=3.12 
(139 d.o.f), as also evident from the data and residuals shown in Fig. 6. Similar results 
are also obtained for the 2000 observations. 
As we discussed in sect. 2, this high $\chi^2$ is due to the large deviation 
between 0.3 and 0.4 keV, corresponding to the carbon edge feature in the detector effective 
area (Orr et al. 1997). 
Moreover, it is possible to see how this feature affects the LECS continuum over a 
quite large energy range.
When the energy bins below 0.8 keV are not included in the best fit the reduced
$\chi^2$ decreases down to much better value of 1.25 (105 d.o.f.). As stated in Sect.~2, 
we tried to adjust the instrumental parameter to reduce the inconvenience due to this 
spurious features. We applied the XSPEC command GAIN to the LECS data correcting the 
energy channel relation with a linear law and found that the feature practically 
disappeared as shown in the lower panel of Fig.~6. 
To be sure that this correction
does not alter the estimate of the spectral parameters, we evaluated their best fit
values after eliminating the energy bins between 0.1 and 0.8 keV and they generally remained
unchanged within the statistical errors.

Large $\chi^2$ values are also obtained from the 1999--2000 MECS data. An inspection of the residuals 
shows that they are likely due to systematics in the response matrices of the instrument, 
which are much larger than statistical uncertainties because of the very high number
of events. The accuracy of calibration of this instrument is indeed of about 2\%
(Fiore et al. 1999), and we verified that it is sufficient to take into account a 
systematic error of 1.5\% to reduce the $\chi^2$ to well acceptable values.

We therefore decided to apply these corrections in the spectral analysis of all high
statistics observations (1999-2000) and limited the useful LECS range to 0.1--2 keV.

\subsection{The 1999 observation}

During this observation the mean count rate level of Mkn~421 was intermediate between
those measured in 1997 and 1998. The flux in the PDS band was therefore not high and these
data were used in the spectral fit of the whole observation only up to 25 keV. Although
the used range was narrower than in the 1997--1998, we detected a significant
spectral curvature, with an average $b$=0.42, very similar to the results of 1997.
The value of $a$ is higher and determines a lower value of $E_p$. 
The resulting SED, plotted in Fig. 7, is similar to those of the previous observations.

We evaluated the log-parabolic best fits also in four shorter time intervals to study 
the possible evolution of the SED. These intervals, identified by the roman numerals
from I to IV, correspond to the following time window in the light curve of Fig. 1:
I from 0 to 62,000 s, II from 62,000 to 190,000 s, III from 190,000 to 245,000 s and
IV from 245,000 to 290,000 s. In particular interval II corresponds to the rather stable
portion of the light curve, interval III corresponds to the central portion of the flare 
while the other two are the initial and final segments of the light curve.
We had a large number of events in each of these intervals: the lowest number of
MECS counts was about 35,000 (interval IV), but we chose to exclude PDS data 
because of the low net signal in each interval.
The resulting reduced $\chi^2$ values, given in Table 4, are fully satisfactory and 
the spectral parameters are given in Table 5; the corresponding SEDs in the LECS and MECS 
energy bands are plotted in Fig. 8 with their best fits. 
Note that the curvature parameter $b$ changed very little from the 
mean value for the entire data set, whereas the normalisation $K$ changed by a 
factor of $\sim$2.
This result implies that the spectral changes were rather limited during 
the pointing, and in particular, the spectral curvature during the flare was the same
measured in the other intervals.

\subsection{The April 2000 observation}
In spring 2000 Mkn~421 reached the highest brightness level of all the {\it Beppo}SAX 
observations. The source was well detectable in the PDS at energies
higher than 100 keV and therefore we considered in our analysis the PDS 13--120 keV data.

The resulting $\chi^2$ for the entire observation (see Table 4) was well acceptable 
and the corresponding log-parabolic SED in shown in Fig.~7. 
The source spectrum in this high state was very different from that observed
in fainter states. In particular, the spectral curvature $b$ decreased to 0.21, and the
peak energy increased at about 3 keV.

From the light curve of Fig. 2 we see that the X-ray flux of Mkn~421 was characterised
by an approximately oscillating behaviour with a typical amplitude of about 1.5, without
prominent flares. We selected then two data sets, useful for a more detailed analysis,
on the base of the count rate. All the events in the time intervals during which the 
mean 2--10 keV MECS count rate was higher than 7.5 ct/s were included in the High state subset
{\bf H}, while the remaining events were in the Faint subset {\bf F}. The parameters of the 
log-parabolic best fits of these two subsets are given in Table 5 and the corresponding SEDs
are plotted in Fig. 9. 

Again spectral changes between the two states are rather small: 
in particular the value of $b$ was practically unchanged  while $a$ changed by 0.08, implying 
an increase of the peak energy from 2.3 to 3.7 keV in the {\bf H} state.

\begin{figure}
      \vspace{0.5cm}
\epsfysize=9cm
\epsfbox{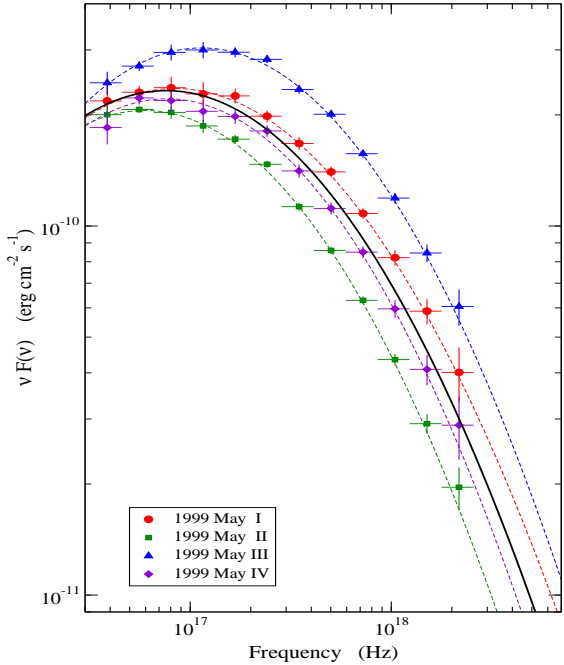}
\caption[]{
The Spectral Energy Distributions of Mkn~421 in the four time segments of the May 1999
observation. Dashed lines are the log-parabolic best fits, while the thick solid line 
is the best fit of the entire observation.
}
\end{figure}

\begin{figure}
      \vspace{0.5cm}
\epsfysize=9cm
\epsfbox{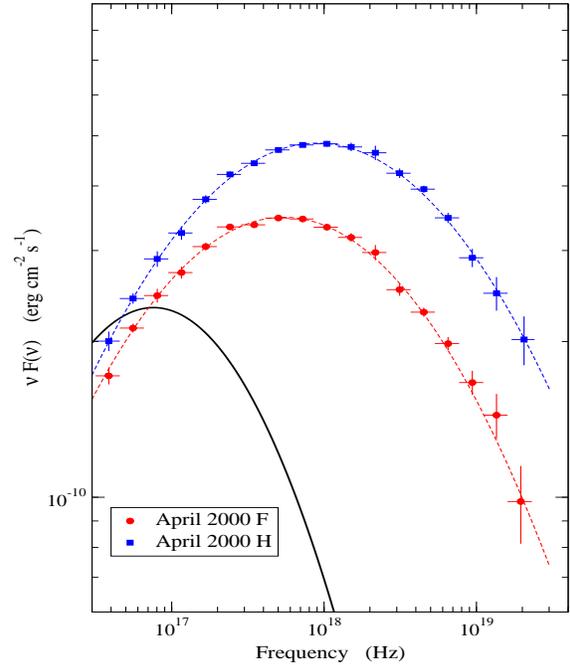}
\caption[]{
The Spectral Energy Distributions of Mkn~421 in the High and Faint states during the April 2000
observation. Dashed lines are the log-parabolic best fits, the thick solid line is the best fit 
of the entire May 1999 observation, the same plotted in Fig. 8, shown for a better comparison 
among the different luminosity states of the source.
}
\end{figure}

\subsection{The May 2000 observation}

During the last observation of May 2000 Mkn~421 was in a high luminosity state.
The spectral best fit for the entire data set was evaluated using the same energy ranges 
for LECS and MECS of April 2000 observation, while that of PDS was limited at 70 keV.
The resulting reduced $\chi^2$ is reported in Table 4. The mean spectral continuum is very well 
described by the log-parabolic law, as shown by the SED plotted in Fig.~7.
\begin{table*}
\caption{ Best fit spectral parameters of the log-parabolic model 
for the 1999 and 2000  observations of Mkn~421.}
\label{tab1}
\begin{tabular}{lccccccc}
\hline
Date & $a$ & $b$ & $K$ & $E_p (keV)$  & $\nu_p F(\nu_p)^{(2)}$ & $F_{bol}^{(2)}$ & $F_{2-10~keV}^{(2)}$\\
\hline
1999-05-04/08       & 2.421 (.004) & 0.419 (.005) & 1.140 (.007)~~10$^{-1}$ & 0.315 (.006) & 2.33 (.02) & ~9.7 (.1)            & 1.10 (.01)\\ 
1999-05 I$^{(1)}$   & 2.36~~ (.01)~ & 0.39~~  (.01)~  & 1.22~~~  (.02)~~10$^{-1}$  & 0.35~~ (.01)~  & 2.37 (.05) & 10.2 (.2)   & 1.31 (.03)\\ 
1999-05 II$^{(1)}$  & 2.539 (.006) & 0.442 (.008) & 0.88~~~  (.01)~~10$^{-1}$  & 0.25~~  (.01)~  & 2.06 (.03) & ~8.3 (.2)      & 0.72 (.02)\\ 
1999-05 III$^{(1)}$ & 2.294 (.007) & 0.450 (.009) & 1.70~~~  (.02)~~10$^{-1}$  & 0.47~~  (.01)~  & 3.04 (.04) & 12.2 (.2)      & 1.89 (.03)\\ 
1999-05 IV$^{(1)}$  & 2.45~~ (.01)~  & 0.45~~  (.01)~  & 1.07~~~  (.02)~~10$^{-1}$  & 0.32~~  (.01)~  & 2.22 (.05) & ~8.9 (.2) & 0.98 (.02)\\ 
 &  &  &  &  &  &                                                                                                              &  \\  
2000-04-26/30 & 1.805 (.002) & 0.212 (.002) & 2.270 (.007)~~10$^{-1}$ & 2.88 (.04)  & 4.03 (.01) & 23.5 (.1)                   & 6.25 (.03)\\  
2000-04 H     & 1.765 (.004) & 0.205 (.003) & 2.59~~~  (.01)~~10$^{-1}$  & 3.7~~  (.1)~   & 4.85 (.03) & 28.8 (.3)             & 7.65 (.08)\\  
2000-04 F     & 1.843 (.003) & 0.222 (.003) & 2.041 (.008)~~10$^{-1}$ & 2.26 (.04)  & 3.49 (.02) & 19.9 (.2)                   & 5.27 (.05)\\  
 &  &  &  &  &  &                                                                                                              &  \\ 
2000-05-09/12 & 1.882 (.003) & 0.180 (.003) & 1.945 (.006)~~10$^{-1}$ & 2.13 (.05) & 3.26 (.01) & 20.6 (.2)                    & 4.95 (.05)\\ 
2000-05 I     & 1.962 (.003) & 0.195 (.003) & 1.734 (.007)~~10$^{-1}$ & 1.25 (.02) & 2.79 (.01) & 17.0 (.1) 		       & 3.86 (.02)\\ 
2000-05 II    & 1.704 (.009) & 0.200 (.007) & 2.87~~~ (.03)~~10$^{-1}$  & 5.5~~  (.4)~ & 5.9~~  (.1)~  & 35.6 (.9)             & 9.32 (.24)\\ 
2000-05 III   & 1.737 (.008) & 0.197 (.007) & 2.62~~~ (.02)~~10$^{-1}$  & 4.6~~   (.3)~ & 5.1~~  (.1)~  & 31.1 (.7)            & 8.12 (.18)\\ 
2000-05 IV    & 1.72~~ (.01)~ & 0.194 (.009) & 2.81~~~ (.03)~~10$^{-1}$  & 5.3~~   (.5)~ & 5.7~~  (.1)~  & 34.7 (.9)           & 8.97 (.23)\\ 
2000-05 V     & 1.88~~ (.01)~ & 0.19~~ (.01)~ & 1.71~~~ (.02)~~10$^{-1}$  & 2.1~~   (.1)~ & 2.86 (.04) & 17.7 (.5)             & 4.35 (.13)\\ 
\hline
\multicolumn{7}{c} { }
\end{tabular}

Errors are given at 1 sigma for one interesting parameter. \\
(1) ~~PDS data not included. \\
(2) ~~In units of 10$^{-10}$ erg cm$^{-2}$ s$^{-1}$. \\
\end{table*}

\begin{figure}
      \vspace{0.5cm}
\epsfysize=9cm
\epsfbox{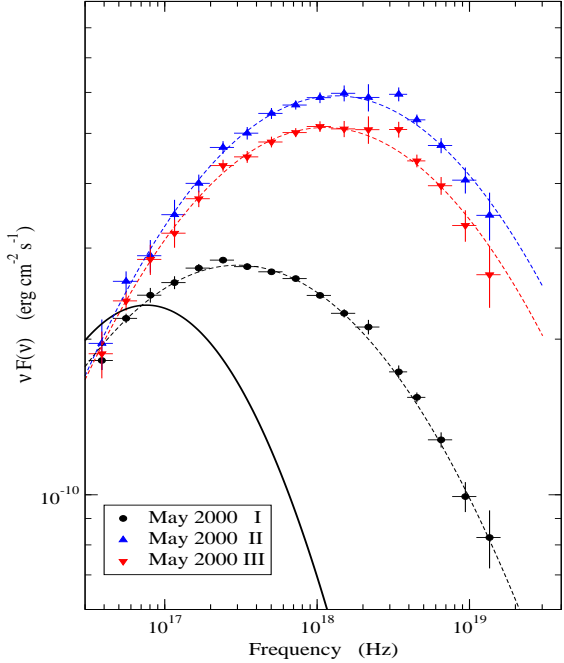}
\caption[]{
The Spectral Energy Distributions of Mkn~421 in the first three time segments of the May 2000
observation. Dashed lines are the log-parabolic best fits. The thick solid line is the best fit 
of the entire May 1999 observation, the same plotted in Fig. 8, shown for a better comparison 
among the different luminosity states of the source.
}
\end{figure}

Also for this long pointing the spectral analysis was performed in five shorter segments,
selected on the basis of the time features in the light curve of Fig. 3. Segment I, from
the beginning to 180,000 seconds, includes the first portion of the light curve before the 
flare; segments II and III were taken in the rising and decaying portions of the flare;
segment IV from 225,000 to 250,000 seconds contains the events of the second rapid flare;
finally, the last segment V includes the remaining data to the end of the observation. 
Log-parabolic best fits, as expected, are generally good (see the reduced $\chi^2$ values 
in Table 4) and systematically much lower than those obtained with the PL+EC model. 
The resulting spectral parameters are also given in Table 5, while the SEDs of the segments 
I, II, III are shown in Fig. 10. 
Those of the two remaining segments are very similar to them -- in particular
that of IV is intermediate between II and III and that of V is close to that of I -- and
have not been plotted to for simplifying the figure. 
We found that the most stable parameter was $b$, remaining within the small range 0.19--0.20, while 
for the entire observation $b$ was lower and equal to 0.18. This apparent inconsistency is likely due 
to the superposition of different spectra with similar curvature but different peak energies 
(varying by a factor of $\sim 4$, see Fig.~10) which makes the overall spectrum appear less curved. 
This effect is indeed not apparent in the 1999 and April 2000 observations when the peak energies 
were more stable.

These results indicate that one of the most relevant change with respect to the 1997--1999 
data is this large decrease of the spectral
curvature, likely associated with the much higher luminosity of the source. However, note
that $b$ did not change greatly on the time scale of weeks from April to May 2000.

\section{Statistical particle acceleration and log-parabolic spectra}
In the previous Sections we have shown that the log-parabolic law is useful to
model the rather mild spectral curvature observed in the X-ray emission of Mkn~421
and in other BL Lac sources. 
It is important to understand if this law is a simple mathematical tool to describe curved 
shapes or if it can be explained by means of some physical process.
In this Section we present some phenomenological considerations on the particle 
spectra from statistical acceleration. We will show that it is possible to obtain 
an integral energy distribution for the particles that follows a log-parabolic law
under quite reasonable hypotheses and derive some simple relations between
the spectral parameters and the most relevant factors affecting the acceleration.
In the subsection 6.1 we will give only a general trace and not a theory
of energy dependent acceleration. More detailed analytical and numerical calculations 
are necessary to confirm the main findings of this approach, and these mathematical 
developments will be the subject of forthcoming works.

\subsection{Energy distribution of accelerated particles}

The energy spectrum of accelerated particles by some statistical mechanism, like
that occurring in a shock wave or in a strong perturbation moving down a jet, 
is usually written as a power law:
\begin{equation} 
N(>\gamma) = N_0 (\gamma/\gamma_0)^{-s+1}  ~~~~~~~, 
\end{equation}
where $N(>\gamma)$ is the number of particles having a Lorentz factor greater than 
$\gamma$  and  $s$  is the spectral index given by:
\begin{equation}
s  = - \frac{Log~p}{Log~\varepsilon}+1   ~~~~~~~~~~~,
\end{equation}
here $p$ is the probability that a particle undergoes an acceleration step 
$i$ in which it has an energy gain equal to $\varepsilon$, generally assumed both 
independent of energy :
\begin{equation}
\gamma_i = \varepsilon \gamma_{i-1}   
\end{equation}
and
\begin{equation}
N_i =  p N_{i-1} = N_0~p^i ~~~~~~~~.
\end{equation}

A log-parabolic energy spectrum follows when the condition that $p$ is independent
of energy is released and one assumes that it can be described by a power relation
as:
\begin{equation}
p_i = g/\gamma_i^q ~~~~~~~~,
\end{equation}
where $g$ and $q$ are positive constants; in particular, for $q>0$ the probability 
for a particle to be accelerated is lower when its energy increases. 
Such a situation can occur, for instance, when particles are confined by a 
magnetic field with a confinement efficiency decreasing for an increasing gyration 
radius. After simple calculations one finds:
\begin{equation}
N_i = N_0~ \frac{g^i} {\prod_{j=0}^{i-1}\gamma_{j}^q} ~~~~~~~~.
\end{equation}
Using Eq.~(12) one can write the product on the right side as:
\begin{equation}
\prod_{j=0}^{i-1}\gamma_{j}^q=\gamma_{0}^{iq} \prod_{j=1}^{i-1}~\varepsilon^{jq}
 =\gamma_{0}^{iq}~ (\varepsilon^{q})^{i(i-1)/2} ~~~~~~~~,
\end{equation} 
where $\gamma_0$ is the initial Lorentz factor of the particles; inserting this result
into Eq.~(15) we obtain:
\begin{equation}
N_i = N_0~ \left(\frac{g}{\gamma_0^q}\right)^i ~(\varepsilon^{q})^{-i(i-1)/2} ~~~~~~~~.
\end{equation}

Finally, combining this equation with Eq.~(12) one can obtain the integral energy 
distribution of the accelerated particles, which is a log-parabolic law:
\begin{equation}
N(>\gamma) = N_0 (\gamma/\gamma_0)^{-[s - 1 + r Log (\gamma/\gamma_0)]} ~~~~~~~,
\end{equation}
with
\begin{equation} 
s = - \frac{Log(g/\gamma_0)}{Log~\varepsilon} - \frac{q-2}{2} 
\end{equation}
and
\begin{equation}
r = \frac{q}{2~Log~\varepsilon} ~~~~~~ .
\end{equation}
The differential spectrum $N(\gamma)$ is:
\begin{eqnarray}
N(\gamma)&=&\frac{N}{\gamma}\frac{d~LogN}{d~Log(\gamma/\gamma_0)} \\
&=&\frac{N_0}{\gamma_0} |s-1+2rLog(\gamma/\gamma_0)|~(\gamma/\gamma_0)^{-s-r Log(\gamma/\gamma_0)}
\nonumber 
\end{eqnarray}
This is not a log-parabolic law, but differs from it very little, because of a factor
only logarithmic dependent on the particle energy. 
Numerical calculations show that the differences 
between this law and a log-parabolic one is smaller than 10\% over a several 
decade wide energy range. Practically, the spectral curvature corresponding to
Eq.~(21) is very similar to that of a log-parabola and cannot be distinguished in
a spectral analysis like that we presented in previous Sections. 
In the following we will assume, therefore, that the differential energy distribution 
of accelerated particles is well approximated by a log-parabola like that results 
from approximating the log term  as a constant which can be 
included in the normalisation.

It is important to note that the spectral parameters given by Eqs.~(19) and (20)
are in a linear relationship; in fact, after eliminating $Log~\varepsilon$ one obtains:
\begin{equation} 
s = - r \left(\frac{2}{q} Log(g/\gamma_0)\right) - \frac{q-2}{2} 
\end{equation}

The assumption of Eq.~(14) about the energy dependence of the acceleration probability
can been modified to take into account other escape processes. For instance, one
can assume that the acceleration probability is constant for low energies and begins
to decrease above a critical Lorentz factor $\gamma_c$; a phenomenological formula
for this can be:
\begin{equation}
p_i = \frac{g}{1 + (\gamma_i / \gamma_c)^q} ~~~~~~~~.
\end{equation}
The asymptotic behaviour of the particle energy distribution derived from such an 
assumption can easily be devised: for $\gamma \ll \gamma_c$ it will follow a power law 
with spectral index $s  = -(Log~g)/(Log~\varepsilon)$, while for $\gamma \gg \gamma_c$ 
it will approximate a log-parabolic spectrum like Eq.~(18). We expect that in this case 
the spectrum of emitted radiation can be described by Eq.~(9).

\subsection{Synchrotron radiation spectrum}

It is important to know the relations between the spectral parameters $a$ and $b$ 
of the emitted radiation and those of the electron population, namely $s$ and $r$.  
The spectral distribution of the SR emitted by relativistic
electrons with a log-parabolic energy distribution cannot be computed analytically
in a simple way, however, for our aim the relations between the spectral indices 
can be derived under the usual $\delta$--approximation and the assumption 
that the electrons are isotropically distributed in a homogeneous randomly oriented 
magnetic field: 
\begin{equation}
P_S(\nu)=\int P(\nu(\gamma)) N(\gamma)~ d\gamma 
\end{equation}
where the power radiated by a single particle is:
\begin{equation}
P(\nu)=C_1 \gamma^2 B^2  ~\delta(\nu - \nu_S) 
\end{equation}
and
\begin{equation}
\nu_S~=~C_2  \gamma^2 B = \gamma^2 \nu_0
\end{equation}
where $C_1$=$(4e^4)/(9m^2c^3)$, $C_2$=$0.29(3e)/(4\pi mc)$ (Rybicki \& Lightman 1979).

From Eq.~(20) then follows:
\begin{equation}
P_S(\nu) \propto (\nu/\nu_0)^{-(a + b Log(\nu/\nu_0))}            
\end{equation}
with
\begin{eqnarray}
 a~&=&~(s-1)/2  \nonumber \\  
 b~&=&~r/4 ~~~~~~.
\end{eqnarray}

It is important to stress that the parameter $a$, as defined above, is different from
that we defined in Eq.~(3): the former being the energy index at the frequency $\nu_0$,
whereas the latter the photon index at the energy $E_1$, that in our analysis has been 
fixed at 1 keV, not corresponding to the frequency $\nu_0$.
It is not difficult to verify that these two parameters differ for an additional 
constant, whose value depends on $\nu_0$ and $\nu_1=E_1/h$. 
We stress, however, that Eq.~(22) implies that also $a$ and $b$ are expected to be 
{\it linearly dependent} on each other.

\section{Discussion}
The interpretation of the spectral variability of BL Lac objects is not a simple
problem because it involves the description of non stable processes with many
unknown physical quantities. It is important, therefore, in the spectral 
data analysis, to use models characterised by rather simple analytical formulae
which can be directly related to some physical parameters of the source.
In our analysis of the {\it Beppo}SAX wide band observations of Mkn~421, covering in some
occasions about three decades in energy, we used a log-parabolic spectral model 
which represents a further step in this direction.
We have shown that the curved spectra of these sources are very well represented
by a law of this type, as opposed to power laws with exponential cut-off for which 
we did not obtain generally acceptable fits.

As mentioned in the Introduction the fact that the mild curvature of blazar spectra can
be well described by a log-parabolic law was noticed for the first time by Landau
et al. (1986), but these authors did not give a satisfactory model justification 
for it.
A widely accepted interpretation of the spectral curvature is in terms of radiation 
cooling of high energy electron population, injected with a power law spectrum, via 
synchrotron and inverse Compton processes. 
In rapidly variable sources, like blazars, there are several processes and time scales which
compete in determining the observed light curves and spectral shapes (see, for
instance, Massaro et al. 1996): high energy particle acceleration, injection and 
diffusion through the emitting volume, radiative cooling, radiation crossing time
and particle escape are all contributing processes which may be associated with different time scales and 
it is not simple to establish which is the dominant one that determines the observed variability.
In the case of HBL sources, in which the synchrotron emission peaks 
in the X rays the radiative cooling time can be very short:
\begin{eqnarray}
 t_{rad}~&=&~~3.39\times10^{10} (B^{1/2}/u)~ \nu^{-1/2}~ \nonumber \\
 &<& 8.52\times10^{11}~B^{-3/2}~ \nu^{-1/2} ~~{\rm s} ~~~,
\end{eqnarray}   
where $u = (B^2/8\pi) + w_{ph}$ is the energy density of the magnetic and photon
fields ($B$ in Gauss), the upper limit in Eq.~(29) corresponds to neglecting the photon 
energy density. 

\begin{figure}
\epsfysize=9cm
\epsfbox{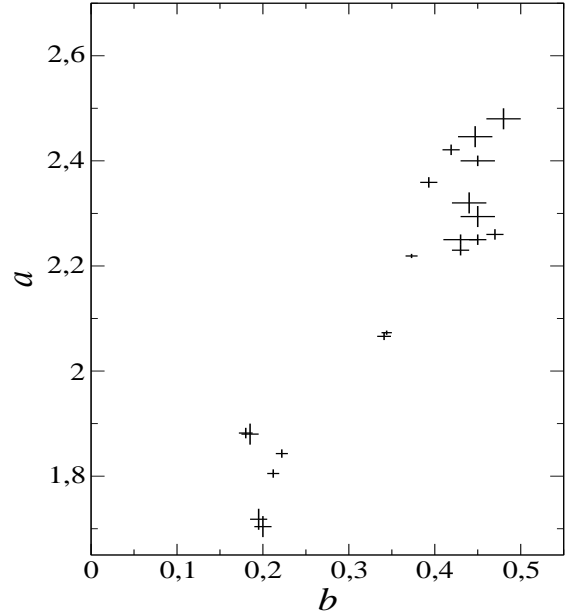}
\caption[]{
The high correlation between the two spectral parameters $a$ and $b$ of the log-parabolic
law supports the energy dependent statistical acceleration of the emitting electrons.
}
\end{figure}

In the bulk frame of the electrons moving down the jet, for frequencies 
$\nu \simeq 10^{16}$ Hz (we assume a beaming factor of about 10) and an 
effective magnetic field of $\sim$ 1 Gauss, we obtain radiative lifetimes of the 
order of one hour, corresponding to minutes in the observer's frame. 
The emission can be maintained only if the acceleration mechanisms are continuously 
at work.

The most common approach to produce curved spectra is to consider radiative losses 
and the escape of high energy electrons from the emitting region: the literature
has been summarized, for the other TeV blazar Mkn~501, in the recent paper by 
Krawczynski et al. (2002). The intrinsic difficulty is that numerical calculations 
are necessary to solve the transfer equation of the electrons and consequently it 
is hard to introduce a parameter for describing the spectral curvature and to find 
how it is related to the other main physical parameters of the model.

We followed a quite different approach, based on the hypothesis that the
particle spectrum is curved from injection. In Sect.~6 we have shown that 
log-parabolic spectra are naturally obtained for an energy dependent probability 
of statistical acceleration and derived how the curvature parameter $b$ can
be simply related to the acceleration gain $\varepsilon$.
The fact that the spectra of the synchrotron component of several BL Lac objects
have a log-parabolic shape with a rather narrow range of $b$ values (see, for 
instance the SED catalog by Giommi et al. 2002) strongly suggests that it can be a 
``general'' characteristic of these sources. 
It is unlikely that acceleration, loss and escape processes combine to produce such 
similar spectra.

In a simplified dimensional approach the transfer equation can be written as:
\begin{equation}
\frac{N}{t} + \frac{N}{t_{rad}(\gamma)} = Q(\gamma)
\end{equation}
in which only the radiative cooling has been considered and where $Q(\gamma)$ is
the source function. In the limit of a very long cooling we obtain 
$N(\gamma,t) \simeq Q(\gamma)~t$, while for a very fast cooling one has 
$N(\gamma,t) \simeq Q(\gamma)~t_{rad} \propto Q(\gamma)/\gamma$.
We see, therefore, that in the case of an injected spectrum given by a log-parabolic
law, the equilibrium particle spectrum has the same
curvature parameter while the constant index $s$ is increased by unity.

An interesting check if the log-parabolic curve is actually related to the statistical
acceleration is the existence of a linear relation between the two spectral 
parameters $a$ and $b$, as a consequence of that given by Eq.~(22). In Fig.~11
we plotted the points corresponding to the best fit values given in  Tables 3 
and 5. It is evident that the two quantities are positively correlated with a 
linear correlation coefficient equal to 0.940.
This result can be considered a further relevant indication supporting the hypothesis
that the curvature is caused by the particle acceleration process. 
High luminosity states are then related to a greater acceleration gain
$\varepsilon$. In our analysis we measured a change of $b$ from $\sim$0.45 to 
$\sim$0.20, their ratio would then be equal to that of the logarithms of the 
acceleration gains in the two states -- see Eqs.~(20), (27). 
We can therefore estimate that in the high state the acceleration gain 
was 
\begin{equation}
\varepsilon_h = \varepsilon_f^{(b_f/b_h)}~~~~, 
\end{equation}
where $f$ and $h$ indicate the faint and high state, respectively.
With the above $b$ values, we have $\varepsilon_h = \varepsilon_f^{2.2}$,
thus for $\varepsilon_f \sim$1.2--1.5, we obtain that in the high state the
acceleration efficiency would increase by about 50\%, and does not require
relevant modifications of the source structure. 

Another important consequence of the log-parabolic modelling of blazar's SEDs 
is that it can be useful to distinguish the possible presence of several emission 
components at the same time.
The optical--X-ray SED of Mkn~421 in the April 1998 observation (Fig. 5)
indicates that simple extrapolations of the X-ray spectra down to the optical 
range does not match nearly simultaneous data. Furthermore, the light curves 
plotted in Figs. 1, 2 and 3 show that the amplitude of the variations is
much higher at energies higher than 2 keV than below. This behaviour is
naturally explained if the emission in the soft X rays is due to the
superposition of two components: one associated with the slowly variable 
optical-UV emission and the other, dominating above $\sim$1 keV, responsible 
for the larger brightness changes.

Disentangling different emission components, on the basis
of SED structure and variability, can be achieved only with coordinated observational  
campaigns covering a very wide spectral range, from IR to high energy $\gamma$ rays.
This is certainly one of the most important scientific targets for future high energy 
astrophysics space observatories (e.g. Swift, AGILE and GLAST) and it is important
to organize this work in advance to obtain very fruitful results. 

\begin{acknowledgements}

This work is based on {\it Beppo}SAX data available from the public archive at the 
ASI Science Data Center. The authors are grateful to F. Fiore for useful suggestion 
on the data analysis and to A. Cavaliere and G. Ghisellini for fruitful discussions 
on the modelling of BL Lac objects.
The CNR Institutes and the {\it Beppo}SAX Science Data Center
are financially supported by the Italian Space Agency (ASI) in the
framework of the {\it Beppo}SAX mission. 
Part of this work was performed with the financial support Italian MIUR 
(Ministero dell' Istruzione Universit\'a e Ricerca) under the grant 
Cofin 2001/028773.

\end{acknowledgements}


\begin{thebibliography}{ }
\small

\bibitem[Boella et al. 1997]{boe} Boella G., Chiappetti L., Conti G. et al. 1997b, A\&AS 122, 327

\bibitem[Fiore et al. 1999]{fiore} Fiore F., Guainazzi M., Grandi P. 1999, Cookbook for 
{\it Beppo}SAX NFI Spectral Analysis (http://www.sdc.asi.it/ \\ software)

\bibitem[Fossati et al 2000a]{fossat1} Fossati G., Celotti A., Chiaberge M. et al. 2000a
ApJ 541, 153

\bibitem[Fossati et al 2000b]{fossat2} Fossati G., Celotti A., Chiaberge M. et al. 2000b
ApJ 541, 166

\bibitem[Fukugita et al. 1995]{fukug}Fukugita M., Shimasaku K., Ichikawa T. 1995,
PASP 107, 945 

\bibitem[Giommi et al. 2002]{giomcat}Giommi P., Capalbi M., Fiocchi M. et al. 2002, Proc.
Blazar Astrophysics with {\it Beppo}SAX and Other Observatories (P. Giommi, E. Massaro,
G.G.C. Palumbo eds.), ASI Special Publication, p. 63

\bibitem[Inoue Takahara 1996]{inotak}Inoue S., Takahara F. 1996, ApJ 97, 1

\bibitem[Krawczynski 2002]{kraw02} Krawczynski H., Coppi P.S., Aharonian F. 2002 MNRAS 336, 721

\bibitem[Krennrich et al 1999]{krennr} Krennrich F., Biller S.D., Bind I.H. et al., 1999
ApJ 511, 149

\bibitem[Landau 1986]{landau} Landau R. et al. 1986, ApJ 308, 78

\bibitem[LokmanSavage 1995]{lokman} Lockman F.J., Savage B.D. 1995, ApJS 97, 1

\bibitem[Malizia 2000]{malizia00} Malizia A., Capalbi M., Fiore F. et al. 2000, MNRAS 312, 123

\bibitem[Massaro 1996]{mass96} Massaro E., Nesci R. et al. 1996, A\&A 314, 87

\bibitem[Massaro 2000]{mass00} Massaro E., Cusumano G. et al. 2000, A\&A 361, 695

\bibitem[Massaro 2003]{mass03} Massaro E., Perri M. et al. 2003, A\&A 399, 33

\bibitem[Mead 1990]{mead90} Mead A.R.G., Ballard K.R., Brand P.W.J.L., et al. 1990, A\&AS 83, 183

\bibitem[Orr 1997]{orr97} Orr A., Parmar A., Guainazzi M., Fiore F., Ricci D. 1997, 
{\it Beppo}SAX Science Data Center Technical Report N. 15

\bibitem[Padovani 1995]{padov} Padovani P., Giommi P. 1995, ApJ 111, 222

\bibitem[Parmar et al. 1997]{par} Parmar A.N., Martin D.D.E., Bavdaz
M. et al., 1997, 
A\&AS 122, 309

\bibitem[Perri 2003]{perri} Perri M., Massaro E., Giommi P. et al. 2003, A\&A 407, 453

\bibitem[Pian et al. 1998]{pian98} Pian E., Vacanti G., Tagliaferri G. et al. 1998, ApJ 
492, L17

\bibitem[Punch et al. 1992]{punch} Punch M. et al. 1992, Nature 160, 477

\bibitem[Rybicki & Lightman 1979]{RybLig} Rybicki G.B., Lightman A.P. 1979, Radiative 
Processes in Astrophysics, Wiley, New York

\bibitem[Tanihata 2002]{tanih02} Tanihata C. et al. 2002, ApJ in press (astro-ph/0210214)

\bibitem[Tosti 1998]{tosti98} Tosti G., Fiorucci M., Luciani M., et al. 1998, A\&A 339, 41

\bibitem[Urry et al. 2000]{urry} Urry C.M., Scarpa R., O'Dowd M. et al. 2000, ApJ 532, 816

\bibitem[Villata et al. 1998]{villat} Villata M., Raiteri C.M., Lanteri L. et al. 1998
A\&AS 130, 305 

\bibitem[Zhang 2002]{zan02} Zhang Y.H. 2002, MNRAS 337, 609

\end{thebibliography}
\end{document}